\documentclass[superscriptaddress,final,preprint,showkeys]{revtex4-1}
\usepackage[dvips]{graphicx}
\usepackage[utf8]{inputenc}
\usepackage{multirow,booktabs}
\usepackage{amsmath,amssymb}

\usepackage[table]{xcolor}
\usepackage{xcolor}
\usepackage[%
  colorlinks=true,
  urlcolor=blue,
  linkcolor=blue,
  citecolor=blue
]{hyperref}

\begin{document}
\title{The dynamical structure of political corruption networks}

\author{Haroldo V. Ribeiro}\email{hvr@dfi.uem.br}
\affiliation{Departamento de F\'isica, Universidade Estadual de Maring\'a, Maring\'a, PR 87020-900, Brazil}

\author{Luiz G. A. Alves}
\affiliation{Institute of Mathematics and Computer Science, University of S\~ao Paulo, S\~ao Carlos, SP 13566-590, Brazil}

\author{Alvaro~F.~Martins}
\affiliation{Departamento de F\'isica, Universidade Estadual de Maring\'a, Maring\'a, PR 87020-900, Brazil}

\author{Ervin K. Lenzi}
\affiliation{Departamento de F\'isica, Universidade Estadual de Ponta Grossa, Ponta Grossa, PR 84030-900, Brazil}

\author{Matja\v{z} Perc}\email{matjaz.perc@uni-mb.si}
\affiliation{Faculty of Natural Sciences and Mathematics, University of Maribor, Koro\v{s}ka cesta 160, SI-2000 Maribor, Slovenia}
\affiliation{CAMTP -- Center for Applied Mathematics and Theoretical Physics, University of Maribor, Mladinska 3, SI-2000 Maribor, Slovenia}
\affiliation{Complexity Science Hub, Josefst{\"a}dterstra{\ss}e 39, A-1080 Vienna, Austria}

\begin{abstract}
Corruptive behaviour in politics limits economic growth, embezzles public funds, and promotes socio-economic inequality in modern democracies. We analyse well-documented political corruption scandals in Brazil over the past 27 years, focusing on the dynamical structure of networks where two individuals are connected if they were involved in the same scandal. Our research reveals that corruption runs in small groups that rarely comprise more than eight people, in networks that have hubs and a modular structure that encompasses more than one corruption scandal. We observe abrupt changes in the size of the largest connected component and in the degree distribution, which are due to the coalescence of different modules when new scandals come to light or when governments change. We show further that the dynamical structure of political corruption networks can be used for successfully predicting partners in future scandals. We discuss the important role of network science in detecting and mitigating political corruption.
\end{abstract}
\maketitle
\linespread{1.1}

\section*{Introduction}
The World Bank estimates that the annual cost of corruption exceeds $5\%$ of the global Gross Domestic Product (US~\$$2.6$ trillion), with US~\$$1$ trillion being paid in bribes around the world~\cite{wb2008,wb2016}. In another estimation, the non-governmental organization Transparency International claims that corrupt officials receive as much as US~\$$40$ billion bribes per year in developing countries~\cite{transparency}. The same study also reports that nearly two out of five business executives had to pay bribes when dealing with public institutions. Despite the difficulties in trying to estimate the cost of global corruption, there is a consensus that massive financial resources are lost every year to this cause, leading to devastating consequences for companies, countries, and society as a whole. In fact, corruption is considered as one of the main factors that limit economic growth~\cite{rose1975economics,shleifer1993corruption,mauro1995corruption,bardhan1997corruption,shao2007quantitative}, decrease the returns of public investments~\cite{haque2008public}, and promote socioeconomic inequality~\cite{mauro1995corruption,gupta2002does} in modern democracies.

Corruption is a long-standing problem in human societies, and the search for a better understanding of the processes involved has attracted the attention of scientists from a wide range of disciplines. There are studies about corruption in education systems~\cite{petrov2004corruption}, health and welfare systems~\cite{maestad2011informal,hanf2011corruption}, labour unions~\cite{banfield1985corruption}, religions~\cite{sandholtz2000accounting,paldam2001corruption}, judicial system~\cite{mauro1995corruption}, police~\cite{carter1990drug}, and sports~\cite{forster2016global}, to name just some among many other examples~\cite{bac1996corruption,bolgorian2011corruption,duenez2012evolving,pisor2015corruption,verma2015bribe,paulus2015worldwide,gachter2016intrinsic}. These studies make clear that corruption is widely spread over different segments of our societies in countries all over the world. However, corruption is not equally distributed over the globe. According to the 2016--Corruption Perceptions Index~\cite{transparency} (an annual survey carried out by the agency Transparency International), countries in which corruption is more common are located in Africa, Asia, Middle East, South America, and Eastern Europe; while Nordic, North America, and Western Europe countries are usually considered ``clean countries''~\cite{transparency}. Countries can also be hierarchically organized according to the Corruption Perceptions Index, forming a non-trivial clustering structure~\cite{paulus2015worldwide}.

From the above survey of the literature, we note that most existing studies on corruption have an economic perspective, and the corruption process itself is empirically investigated via indexes at a country scale. However, a particular corruption process typically involves a rather small number of people that interact at much finer scales, and indeed much less is known about how these relationships form and evolve over time. Notable exceptions include the work of Baker and Faulkner~\cite{baker1993social} that investigated the social organization of people involved in price-fixing conspiracies, and the work of Reeves-Latour and Morselli~\cite{reeves2016bid} that examines the evolution of a bid-rigging network. Such questions are best addressed in the context of network science~\cite{watts1998collective,watts2004new,barabasi_s99,watts_d_99,albert2002statistical,estrada2012structure,barabasi_16} and complex systems science~\cite{gell1988simplicity} -- two theoretical frameworks that are continuously proving very useful to study various social and economic phenomena and human behaviour~\cite{hidalgo2007product,hidalgo2009building,fu_prsb11,pastor_rmp15,wang_z_pr16,gomez2016explaining,perc_pr17,clauset2015systematic}.

The shortage of studies aimed at understanding the finer details of corruption processes is in considerable part due to the difficulties in finding reliable and representative data about people that are involved~\cite{xu2005criminal}. On the one hand, this is certainly because those that are involved do their best to remain undetected, but also because information that does leak into the public is often spread over different media outlets offering conflicting points of view. In short, lack of information and misinformation~\cite{del2016spreading} both act to prevent in-depth research.

To overcome these challenges, we present here a unique dataset that allows unprecedented insights into political corruption scandals in Brazil that occurred from 1987 to 2014. The dataset provides details of corruption activities of 404 people that were in this time span involved in 65 important and well-documented scandals. In order to provide some perspective, it is worth mentioning that Brazil has been ranked 79th in the Corruption Perceptions Index~\cite{transparency}, which surveyed 176 countries in its 2016 edition. This places Brazil alongside African countries such as Ghana (70th) and Suriname (64th), and way behind its neighbouring countries such as Uruguay (21th) and Chile (24th). Recently, Brazil has made news headlines across the world for its latest corruption scandal named ``\textit{Opera\c{c}\~ao Lava Jato}'' (English: ``Operation Car Wash''). The Federal Public Ministry estimates that this scandal alone involves more than US~\$$12$ billion, with more than US~\$$2$ billion associated just with bribes.

In what follows, we apply time series analysis and network science methods to reveal the dynamical organization of political corruption networks, which in turn reveals fascinating details about individual involvement in particular scandals, and it allows us to predict future corruption partners with useful accuracy. Our results show that the number of people involved in corruption cases is exponentially distributed, and that the time series of the yearly number of people involved in corruption has a correlation function that oscillates with a four-year period. This indicates a strong relationship with the changes in political power due to the four-year election cycle. We create an evolving network representation of people involved in corruption scandals by linking together those that appear in the same political scandal in a given year. We observe exponential degree distributions with plateaus that follow abrupt variations in years associated with important changes in the political powers governing Brazil. By maximizing the modularity of the latest stage of the corruption network, we find statistically significant modular structures that do not coincide with corruption cases but usually merge more than one case. We further classify the nodes according to their within- and between-module connectivity, unveiling different roles that individuals play within the network. We also study how the giant component of the corruption network evolves over time, where we observe abrupt growths in years that are associated with a coalescence-like process between different political scandals. Lastly, we apply several algorithms for predicting missing links in corruption networks. By using a snapshot of the network in a given year, we test the ability of these algorithms to predict missing links that appear in future iterations of the corruption network. Obtained results show that some of these algorithms have a significant predictive power, in that they can correctly predict missing links between individuals in the corruption network. This information could be used effectively in prosecution and mitigation of future corruption scandals.

\section*{Methods}

\subsection*{Data Collection}

The dataset used in this study was compiled from publicly accessible web pages of the most popular Brazilian news magazines and daily newspapers. We have used as base the list of corruption cases provided in the Wikipedia article \textit{List of political corruption scandals in Brazil}~\cite{wikilist}, but since information about all the listed scandals was not available, we have focused on 65 scandals that were well-documented in the Brazilian media, and for which we could find reliable and complete sources of information regarding the people involved. We have also avoided very recent corruption cases for which the legal status is yet undetermined. We have obtained most information from the weekly news magazine \textit{Veja}~\cite{veja} and the daily newspapers \textit{Folha de S\~ao Paulo}~\cite{folha} and \textit{O Estado de S\~ao Paulo}~\cite{estadao}, which are amongst the most influential in Brazil. More than $300$ news articles were consulted and most links to the sources were obtained from the aforementioned Wikipedia article. After manual processing, we have obtained a dataset that contains the names of 404 people that participated in at least one of the 65 corruption scandals, as well as the year each scandal was discovered. We make this dataset available as electronic supplementary material (File S1), but because of legal concerns all names have been anonymized. Figure~\ref{fig:1}A shows a barplot of the number of people involved in each scandal in chronological order.

\subsection*{Legal Considerations}

As with all data concerning illegal activities, ours too has some considerations that deserve mentioning. We have of course done our best to curate the data, to double check facts from all the sources that were available to us, and to adhere to the highest standards of scientific exploration. Despite our best efforts to make the dataset used in this study reliable and bias-free, we note that just having the name cited in a corruption scandal does not guarantee that this particular person was found officially guilty by the Brazilian Justice Department. Judicial proceedings in large political corruption scandals can take years or even decades, and many never reach a final verdict. From the other perspective, it is likely that some people that have been involved in a scandal have successfully avoided detection. Accordingly, we can never be sure that all individuals that have been involved in a corruption scandal have been identified during investigations, and in this sense our data may be incomplete. Unfortunately, the compilation of large-scale data on corruption will always be burdened with such limitations. But since our results are related to general patterns of corruption processes that should prove to be universal beyond the particularities of our dataset, we argue that these considerations have at most a negligible impact.

\section*{Results and Discussion}

\subsection*{Growth Dynamics of the Number of People Involved}

\begin{figure*}[t]
\centering
\includegraphics[width=16cm]{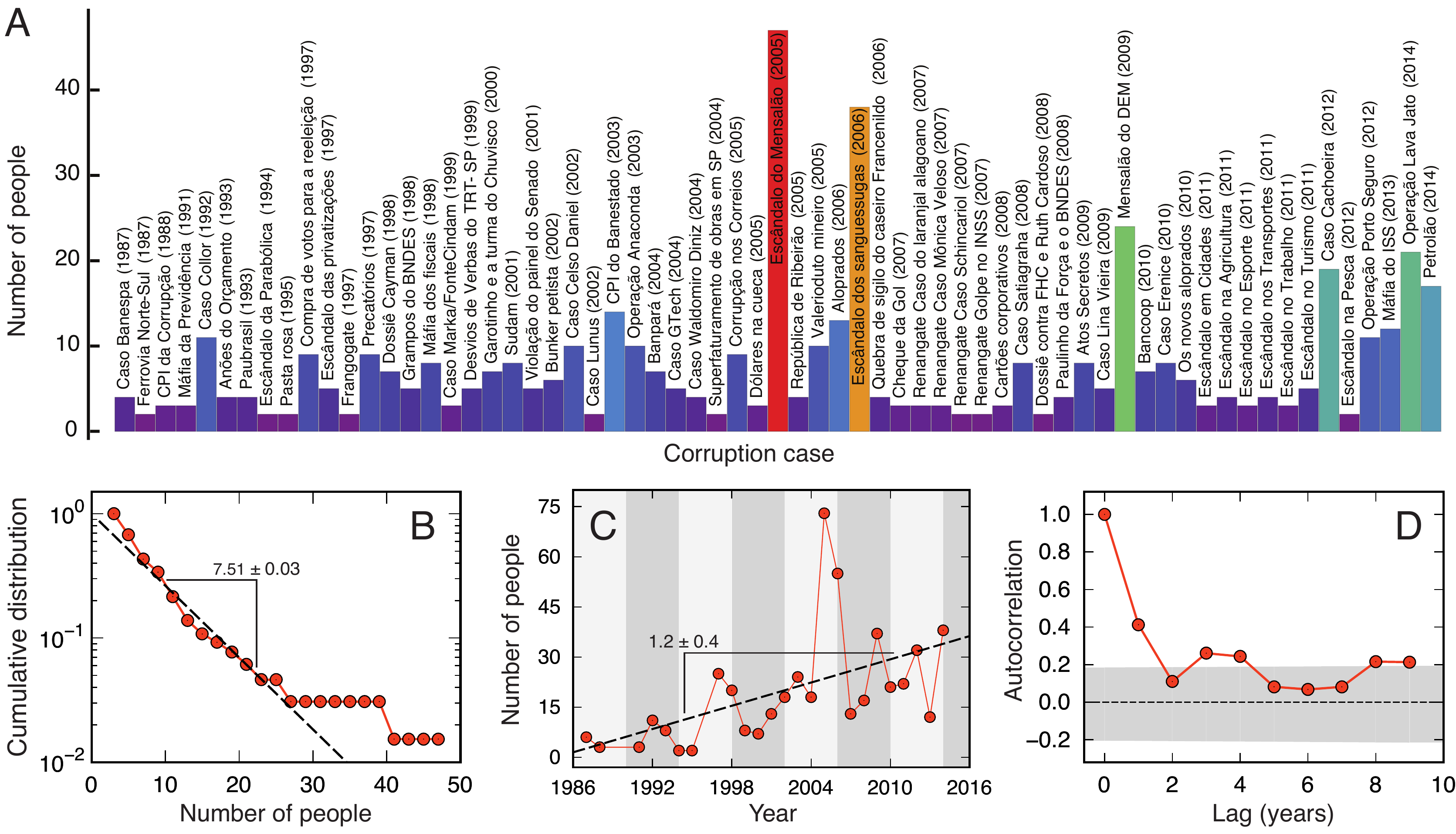}
\caption{Demography and evolving behaviour of corruption scandals in Brazil. A) The number of people involved in each corruption scandal in chronological order (from 1987 to 2014). B) Cumulative probability distribution (on a log-linear scale) of the number of people involved in each corruption scandal (red circles). The dashed line is a maximum likelihood fit of the exponential distribution, where the characteristic number of people is $7.51\pm0.03$. The Cram\'er-von Mises test cannot reject the exponential distribution hypothesis ($p$-value~$=0.05$). C) Time series of the number of people involved in corruption scandals by year (red circles). The dashed line is a linear regression fit, indicating a significant increasing trend of $1.2\pm0.4$ people per year ($t$-statistic~$=3.11$, $p$-value~$=0.0049$). The alternating gray shades indicate the term of each general election that took place in Brazil between 1987 and 2017. D) Autocorrelation function of the time series of the yearly number of people involved in scandals (red circles). The shaded area represents the 95\% confidence band for a random process. It is worth noting that the correlation oscillates while decaying with an approximated four-year period, the same period in which the general elections take place in Brazil.}
\label{fig:1}
\end{figure*}

We start by noticing that the number of people involved in corruption cases does not seem to vary widely (Fig.~\ref{fig:1}A). Having more than ten people in the same corruption scandal is relatively rare (about 17\% of the cases) in our dataset. We investigate the probability distribution of the number of people involved in each scandal, finding out that it can be well described by an exponential distribution with a characteristic number of involved around eight people (Fig.~\ref{fig:1}B). A more rigorous analysis with the Cram\'er-von Mises test confirms that the exponential hypothesis cannot be rejected from our data ($p$-value$\;=0.05$). This exponential distribution confirms that people usually act in small groups when involved in corruption processes, suggesting that large-scale corruption processes are not easy to manage and that people try to maximize the concealment of their crimes.

Another interesting question is related to how the number of people involved in these scandals has evolved over the years. In Fig.~\ref{fig:1}C we show the aggregated number of people involved in corruption cases by year. Despite the fluctuations, we observe a slightly increasing tendency in this number, which can be quantified through linear regression. By fitting a linear model, we find a small but statistically significant increasing trend of $1.2\pm0.4$ people per year ($p$-value$\;=0.0049$). We also ask if this time series has some degree of memory by estimating its autocorrelation function. Results of Fig.~\ref{fig:1}D show that the correlation function decays fast with the lag-time, and no significant correlation is observed for time series elements spaced by two years. However, we further note a harmonic-like behaviour in which the correlation appears to oscillate while decaying. In spite of the small length of this time series, we observe two local maxima in the correlation function that are outside the 95\% confidence region for a completely random process: the first between 3 and 4 years and another between 8 and 9 years. This result indicates that the yearly time series of people involved in scandals has a periodic component with an approximated four-year period, matching the period in which the Brazilian elections take place. Again, this behaviour should be considered much more carefully due to the small length of the time series. Also, this coincidence does not prove any causal relationship between elections and corruption processes, but it is easy to speculate about this parallelism. Several corruption scandals are associated with politicians having exercised improper influence in exchange for undue economic advantages, which are often employed for supporting political parties or political campaigns. Thus, an increase in corrupt activities during election campaigns may be one of the reasons underlying this coincidence.

\subsection*{Network Representation of Corruption Scandals}

\begin{figure*}[!ht]
\centering
\includegraphics[scale=0.55]{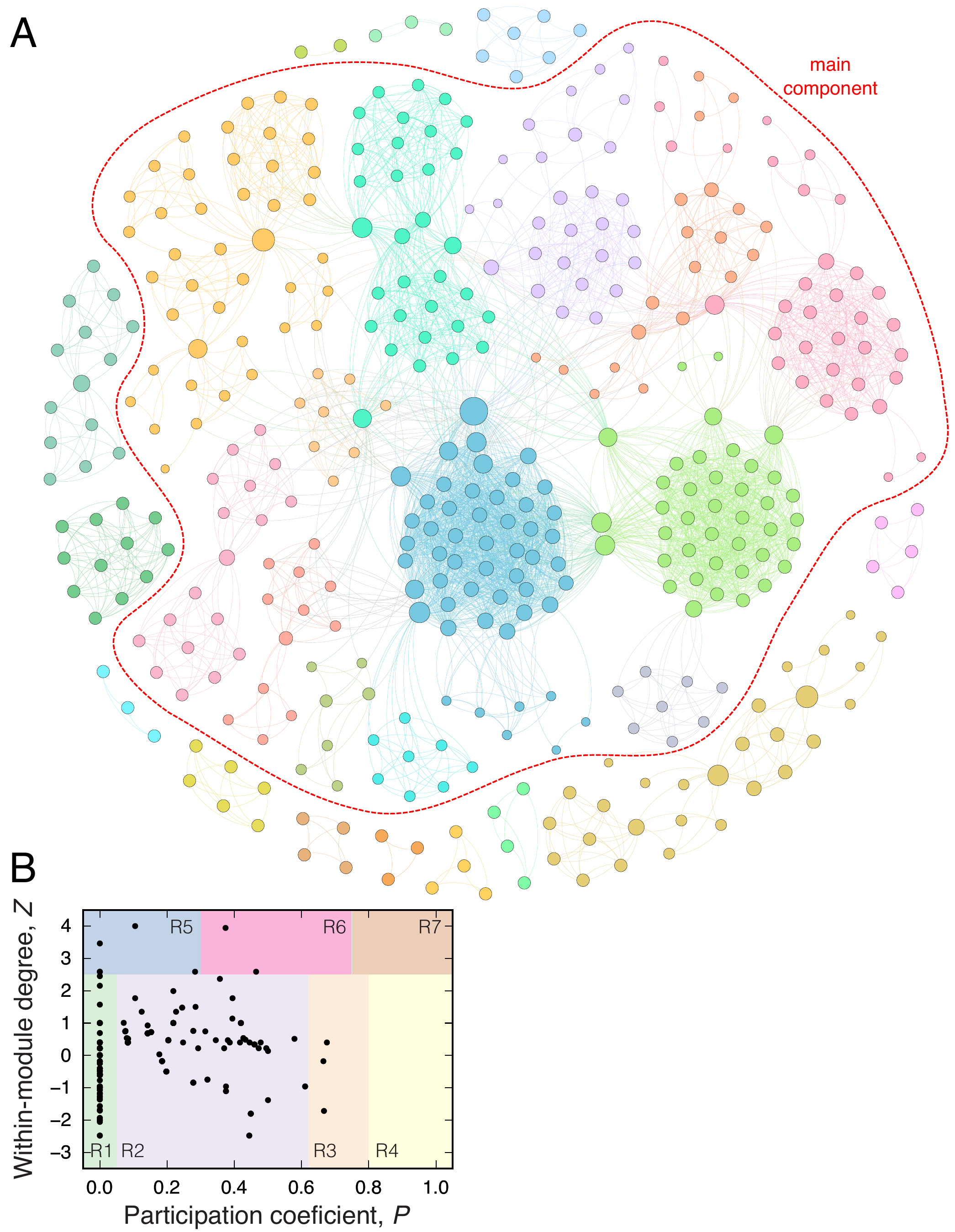}
\caption{Complex network representation of people involved in corruption scandals. A) Complex network of people involved in all corruption cases in our dataset (from 1987 to 2014). Each vertex represents a person and the edges among them occur when two individuals appear (at least once) in the same corruption scandal. Node sizes are proportional to their degrees and the colour code refers to the modular structure of the network (obtained from the network cartography approach~\cite{guimera2005functional,guimera2005cartography}). There are 27 significant modules, and 14 of them are within the giant component (indicated by the red dashed loop). B) Characterization of nodes based on the within-module degree ($Z$) and participation coefficient ($P$). Each dot in the $Z$-$P$ plane corresponds to a person in the network and the different shaded regions indicate the different roles according to the network cartography approach (from R1 to R7). The majority of nodes (97.5\%) are classified as ultraperipheral (R$1$) or peripheral (R$2$), and the remaining are non-hub connectors (R$3$, three nodes), provincial hubs (R$5$, three nodes), and connector hubs (R$6$, two nodes).}
\label{fig:2}
\end{figure*}

In order to study how people involved in these scandals are related, we have built a complex network where nodes are people listed in scandals, and the ties indicate that two people were involved in the same corruption scandal. This is a time-varying network, in the sense that it grows every year with the discovery of new corruption cases. However, we first address static properties of this network by considering all corruption scandals together, that is, the latest available stage of the time-varying network. Figure~\ref{fig:2}A shows a visualization of this network that has 404 nodes and 3549 edges. This network is composed of 14 connected components, with a giant component accounting for 77\% of nodes and 93\% of edges. The average clustering coefficient is $0.925$ for the whole network and $0.929$ for the giant component, values much higher than those expected for random networks with the same number of nodes and edges ($0.044\pm0.001$ for the entire network and $0.069\pm0.002$ for the giant component). This network also exhibits small-world property with an average path length of 2.99 steps for the giant component, a feature that is often observed in most networks~\cite{watts1998collective}. However, this value is greater than the one obtained for a random network ($2.146\pm0.002$), suggesting that in spite of nodes being relatively close to each other, corruption agents have tried to increase their distance as this network has grown up. This result somehow agrees with the so-called ``theory of secret societies'', in which the evolution of illegal networks is assumed to maximize concealment~\cite{baker1993social}. Another interesting property of this network is its homophily (or assortatively), which can be measured by the assortativity coefficient~\cite{newman2002assortative} equal to $0.60$ for the whole network and $0.53$ for the giant component. These values indicate a strong tendency for nodes to connect with nodes of similar degree, a common feature of most social networks~\cite{newman2002assortative} and that here could be related to the growth process of this network. When a new scandal is discovered, all new people are connected to each other and added to the network (so they will start with the same degree). This group of people is then connected to people already present in the network which were also involved in the same scandal. The homophily property suggests that new scandals may act as ``bridges'' between two or more ``important agents'', contributing for making their degrees similar.

Similar to other social networks, this corruption network may also have a modular structure. To investigate this possibility, we have employed the network cartography approach~\cite{guimera2005functional,guimera2005cartography} for extracting statistically significant modules through the maximization of network's modularity~\cite{newman2004finding} via simulated annealing~\cite{kirkpatrick1983optimization}, and also for classifying nodes according to their within- ($Z$, in standard score units) and between-module connectivity (or participation coefficient $P$). This approach yields 27 modules, and 13 of them are just the non-main connected components, and the remaining 14 are within the giant component. The significance of such modules was tested by comparing the network modularity $M$ (the fraction of within-module edges minus the fraction expected by random connections~\cite{guimera2005functional,guimera2005cartography,newman2004finding}) with the average modularity $\langle M_{\text{rand}}\rangle$ of randomized versions of the original network, leading to $M=0.74$ and $\langle M_{\text{rand}}\rangle = 0.201 \pm 0.002$. We note that the number of modules is roughly half of the number of corruption cases (65 scandals), indicating that there are very similar scandals (regarding their components) and suggesting that such cases could be considered as the same scandal.

Network cartography~\cite{guimera2005functional,guimera2005cartography} also provides a classification of the role of nodes based on their location in the $Z$-$P$ plane, as shown in Fig.~\ref{fig:2}B. According to this classification, nodes are first categorized into ``hubs'' ($Z\geq2.5$) and ``non-hubs'' ($Z<2.5$) based on their within-module degree $Z$. Our network has only $7$ hubs and all remaining $397$ nodes are categorized as non-hubs. Nodes that are hubs can be sub-categorized into provincial (R$5$ nodes have most links within their modules, $P<0.3$), connector (R$6$ nodes have about half links within their modules, $0.3\leq P<0.75$), and kinless hubs (R$7$ nodes have fewer than half links within their modules $P\geq 0.75$). The corruption network displays five R$5$ and three R$6$ nodes; there are also two nodes very close to the boundaries R$1$-R$5$ and R$2$-R$6$. Qualitatively, R$5$ people are not so well-connected to other modules when compared with R$6$ people. Non-hubs can be classified into four sub-categories: ultra-peripheral (R$1$ nodes have practically all connections within their modules, $P<0.05$), peripheral (R$2$ nodes have most connections within their modules, $0.05\leq P<0.62$), non-hub connectors (R$3$ nodes have about half links within their modules, $0.62\leq P<0.8$), and non-hub kinless nodes (R$4$ nodes have fewer than 35\% of links within their modules, $P\geq0.8$). In our case, the vast majority of people are classified as R$1$ (198 nodes) and R$2$ (196 nodes), and only three nodes are classified as R$3$ (with two nodes very close to the boundary R$2$-R$3$, Fig.~\ref{fig:2}B). This split is a common feature of biological and technological networks but not of artificial networks such as the Erd\"os-R\'enyi and Barab\'asi-Albert graphs, which generates a much higher fraction of R$3$ and R$4$ (and R$7$ in the Barab\'asi-Albert model). Thus, when a node is not a hub, most connections are within its module, making R$3$ nodes very rare in empirical networks.

\subsection*{Evolution of the Corruption Network}

\begin{figure*}[t]
\centering
\includegraphics[scale=0.5]{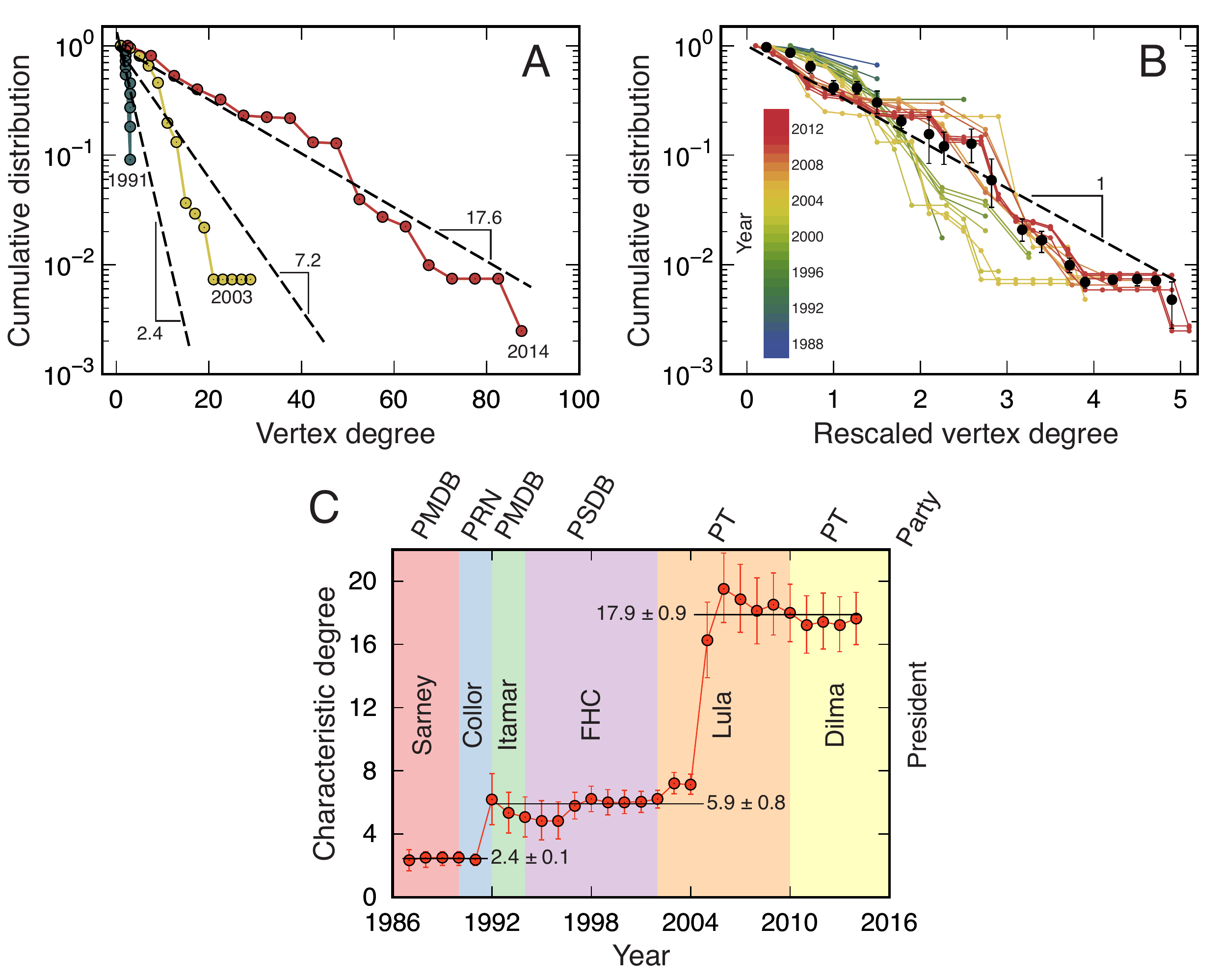}
\caption{The vertex degree distribution is exponential, invariant over time, and the characteristic degree exhibits abrupt changes over the years. A) Cumulative distributions of the vertex degree (on a log-linear scale) for three snapshots of the corruption network in the years 1991 (green), 2003 (yellow), and 2014 (red). The dashed lines are maximum likelihood fits of the exponential distribution, where the characteristic degrees are shown in the plot. We note the widening of these distributions over the years, while the exponential shape (illustrated by the linear behaviour of these distributions on the log-linear scale) seems invariant over time. B) Cumulative distributions of the rescaled vertex degree (that is, the node degree divided by the average degree of the network) for each year of the time-varying corruption network. The colourful points show the results for each of the 28 years (as indicated by the colour code) and the black circles are the window average values over all years (error bars represent 95\% bootstrap confidence intervals). The dashed line is the exponential distribution with a unitary characteristic degree. We note a good quality collapse of these distributions, which reinforces that the exponential distribution is a good approximation for the vertex degree distributions. C) Changes in the characteristic degree over the years. The red circles are the estimated values for the characteristic degree in each year and the error bars stand for 95\% bootstrap confidence intervals. The shaded regions indicate the term of each Brazilian President from 1986 to 2014 (names and parties are shown in the plot). We note that the characteristic degree can be described by three plateaus (indicated by horizontal lines) separated by two abrupt changes: a first between 1991 and 1992 and a second between 2004 and 2005. The first transition is related to the scandal ``\textit{Caso Collor (1992)}'', that led to the impeachment of President Fernando Collor de Mello on corruption charges. The second transition is related with the appearance of five new corruption scandals in 2005, but mainly with the case ``\textit{Esc\^andalo do Mensal\~ao}'', a vote-buying case that involved more than forty people and threatened to topple President Luiz In\'acio Lula da Silva.}
\label{fig:3}
\end{figure*}

Having settled up the basic properties of the latest stage of the time-varying corruption network, we now address some of its dynamical aspects. We start by investigating the time dependence of the degree distribution. Figure~\ref{fig:3}A shows this distribution for the years 1991, 2003 and 2014. We observe the trend of these distributions to become wider over the years, reflecting the growth process of the corruption network caused by the appearance of new corruption scandals. We further note an approximately linear behaviour of these empirical distributions on the log-linear scale, indicating that an exponential distribution is a good model for the degree distribution. By applying the maximum-likelihood method, we fit the exponential model to data and obtain an estimate for the characteristic degree (the only parameter of the exponential distribution). The $p$-values of the Cramer-von Mises test (in chronological order) are $0.03$, $7.6\times10^{-7}$, and $0.01$; thus, the exponential hypothesis can be rejected for the 2003 year. Despite that, the deviations from the exponential model are not very large, which allow us to assume, to a first approximation, that the vertex degree of these networks is exponentially distributed. In order to extend this hypothesis to all years of the corruption network, we have rescaled the degree of each node of the network in a given year by the average degree over the entire network in that particular year. Under the exponential hypothesis, this scaling operation should collapse the different degree distributions for all years onto the same exponential distribution, with a unitary characteristic degree. Figure~\ref{fig:3}B depicts this analysis, where a good collapse of the distributions is observed as well as a good agreement between the window average over all distributions and the exponential model. This result reinforces the exponential model as a first approximation for the degree distribution of the corruption network. The exponential distribution also appears in other criminal networks such as in drug trafficking networks~\cite{wood2017structure} and terrorist networks~\cite{krebs2002mapping}.

Under the exponential hypothesis, an interesting question is whether the characteristic degree exhibits any evolving trend. Figure~\ref{fig:3}C shows the characteristic degree for each year, where we note that its evolution is characterized by the existence of plateaus followed by abrupt changes, producing a staircase-like behaviour that is typically observed in out-of-equilibrium systems~\cite{sethna2001crackling}. Naturally, these abrupt changes are caused by the growth process of the network, that is, by the inclusion of new corruption cases. However, these changes occur only twice and are thus related to the inclusion of special scandals to the corruption network. The first transition occurring between 1991 and 1992 can only be associated with the scandal ``\textit{Caso Collor}'', since it is the only corruption case during this period in our dataset. This scandal led to the \textit{impeachment} of  Fernando Affonso Collor de Mello, President of Brazil between 1990 and 1992. By this time, the network was at a very early stage having only 11 nodes in 1991 and 22 in 1992. Also, when added to the network, people involved in the ``\textit{Caso Collor}'' did not make any other connection with people already present in the network. Thus, the origin of this first abrupt change is somehow trivial; however, it is intriguing that the value of the characteristic degree in 1992 was kept nearly constant up to 2004, when the corruption network was much more developed. The second transition occurring between 2004 and 2005 is more interesting because it involves the inclusion of five new corruption cases (Fig.~\ref{fig:1}A for names), in which people involved were not isolated from previous nodes of the network. As we show later on, the connections of people involved in these five scandals bring together several modules of the network in a coalescence-like process that increased the size of the main component in almost 40\%. Among these five scandals, the case ``\textit{Esc\^andalo do Mensal\~ao}'' stands out because of its size. This was a vote-buying case that involved more than forty people (about twice the number of all other cases in 2005) and threatened to topple the President of Brazil at that time (Luiz In\'acio Lula da Silva). Thus, the second transition in the characteristic degree is mainly associated with this corruption scandal. Another curious aspect of these abrupt changes is that they took place in periods close to important changes in the political power of Brazil: the already-mentioned \textit{impeachment} of President Collor (in 1992) and the election of President Luiz In\'acio Lula da Silva, whose term started in 2003.

\begin{figure*}[!ht]
\centering
\includegraphics[scale=0.5]{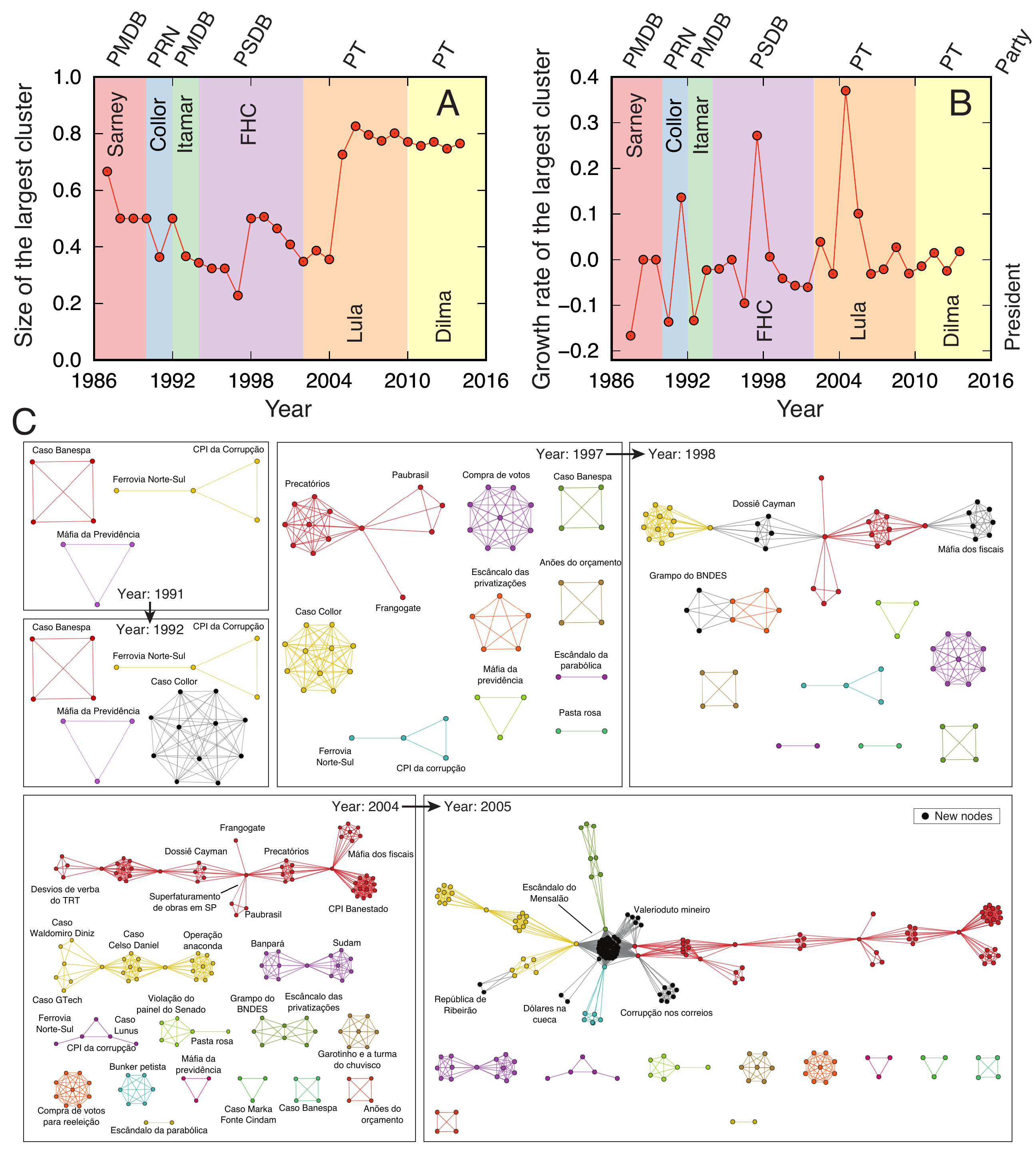}
\caption{Changes in the size of the largest component of the corruption network over time are caused by a coalescence of network modules. A) Evolving behaviour of the fraction of nodes belonging to the main component of the time-varying network (size of the largest cluster) over the years. B) The growth rate of the size of the largest cluster (that is, the derivative of the curve of the previous plot) over the years. In both plots, the shaded regions indicate the term of each Brazilian President from 1986 to 2014 (names and parties are shown in the plot). We note the existence of three abrupt changes between the years 1991-1992, 1997-1998, and 2004-2005. C) Snapshots of the changes in the complex network between the years in which the abrupt changes in the main component took place. We note that between 1991 and 1992, the abrupt change was simply caused by the appearance of the corruption scandal ``\textit{Caso Collor}'', that became the largest component of the network in 1992. The abrupt change between 1997 and 1998 is caused not only by the appearance of three new corruption cases, but also due to the coalescence of two of these new cases with previous network modules. The change between 2004 and 2005 is also caused by the coalescence of previous network components with new corruption cases. In these plots, the modules are coloured with the same colours between consecutive years and new nodes are shown in black. The names of all scandal are shown in the plots.}
\label{fig:4}
\end{figure*}

The changes in the characteristic degree prompt questions about how the corruption network has grown over time. To access this aspect, we study how the size of the main component of the network has changed over time. Figure~\ref{fig:4}A shows the time behaviour of this quantity, where we observe three abrupt changes. These transitions are better visualized by investigating the growth rate of the main component, as shown in Fig.~\ref{fig:4}B. We observe three peaks in the growth rate within the periods 1991-1992, 1997-1998, and 2004-2005. Figure~\ref{fig:4}C shows snapshots of the network before and after each transition, where new nodes and edges associated with these changes are shown in black. It is worth remembering that abrupt changes in 1991-1992 and 2004-2005 were also observed in the characteristic degree analysis, but in 1997-1998 the characteristic degree does not exhibit any change. As previously mentioned, the changes in 1991-1992 just reflect the inclusion of the scandal ``\textit{Caso Collor}''. In 1992, people involved in this corruption case formed a fully connected module that was isolated from the other network components and corresponded to the main component of the network in that year. The other two transitions are more interesting because they involved a more complex process in which people involved in new scandals act as bridges between people already present in the network. In 1998, three new scandals were added to the network and one of them (``\textit{Dossi\^e Cayman}'') established a connection between two modules of the network in 1997. This coalescence-like process has increased the size of the main component from 0.23 to 0.50 (fraction of the nodes). Similarly, four new corruption cases were added to the network in 2005 and one of them (``\textit{Esc\^andalo do Mensal\~ao}'') connected four modules of the network in 2004, increasing the size of the main component from 0.36 to 0.73. We further observe that only a few people already present in the network before the transitions are responsible for forming the modules.

\subsection*{Predicting Missing Links in Corruption Networks}

\begin{figure}[t]
\centering
\includegraphics[width=0.47\textwidth]{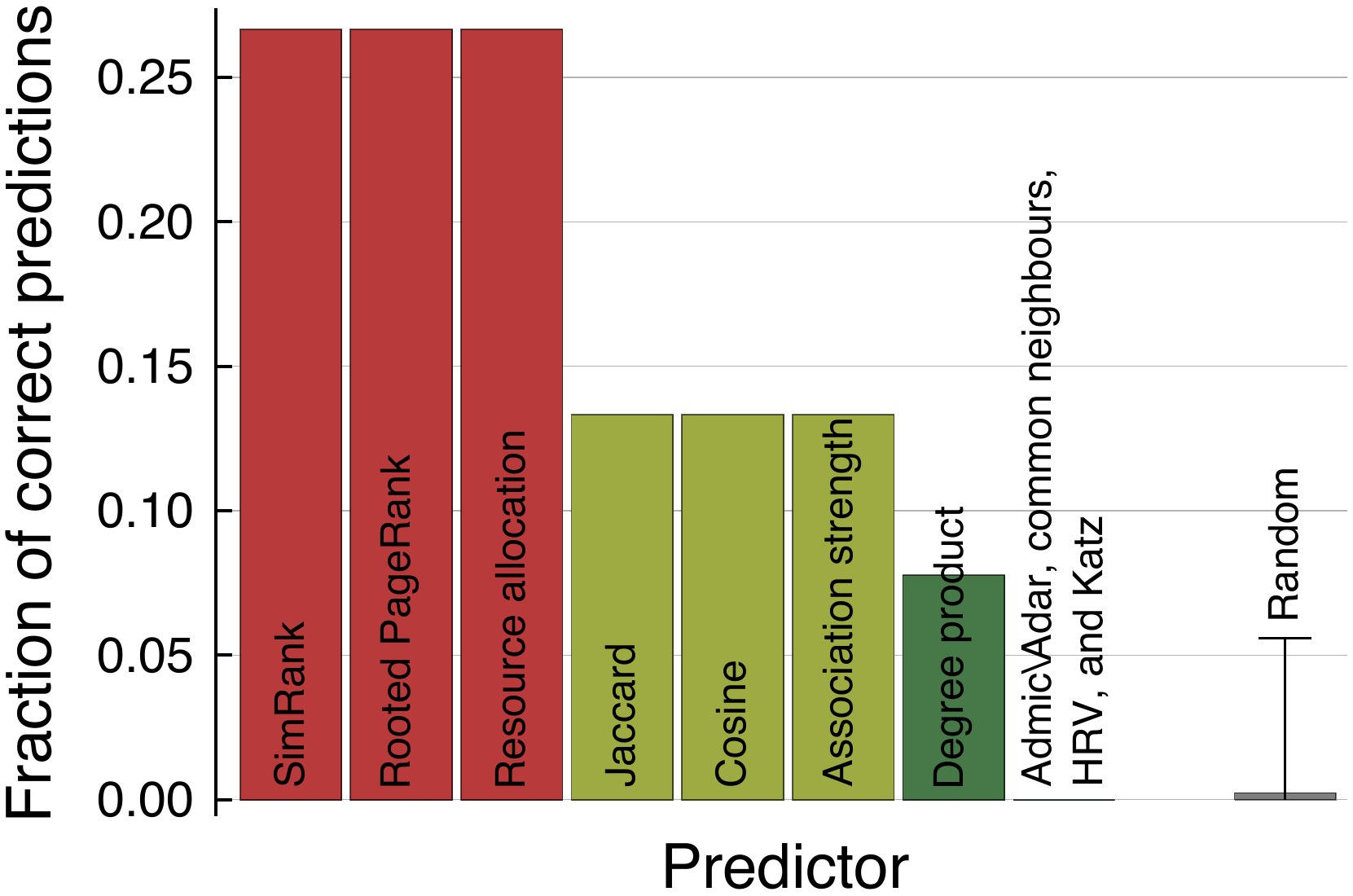}
\caption{Predicting missing links between people in the corruption network may be useful for investigating and mitigating political corruption. We tested the predictive power of eleven methods for predicting missing links in the corruption networks. These methods are based on local similarity measures~\cite{lu2011link} (degree product, association strength, cosine, Jaccard, resource allocation, Adamic-Adar, and common neighbours), global (path- and random walk-based) similarity measures (rooted PageRank and SimRank), and on the hierarchical structure of networks~\cite{clauset2008hierarchical} (hierarchical random graph -- HRV). To access the accuracy of these methods, we applied each algorithm to snapshots of the corruption network in a given year (excluding 2014), ranked the top-10 predictions, and verified whether these predictions appear in future snapshots of the network. The bar plot shows the fraction of correct predictions for each method. We also included the predictions of a random model where missing links are predicted by chance (error bars are 95\% bootstrap confidence intervals).}
\label{fig:5}
\end{figure}

From a practical perspective, the network representation of corruption cases can be used for predicting possible missing links between individuals. Over the past years, several methods have been proposed for accomplishing this task in complex networks~\cite{lu2011link}. In general terms, these methods are based on evaluating similarity measures between nodes, and missing links are suggested when two unconnected nodes have a high degree of similarity. One of the simplest methods of this kind is based on the number of common neighbours, where it is assumed that nodes sharing a large number of neighbours are likely to be connected. There are also methods based on the existence of short paths between nodes. For example, the SimRank method~\cite{jeh2002simrank} measures how soon two random walkers are expected to meet at the same node when starting at two other nodes. Another important approach is the hierarchical random graph~\cite{clauset2008hierarchical} (HRV) method, which is based on the existence of hierarchical structures within the network. Here we have tested 11 of these approaches in terms of their accuracy in predicting missing links in the corruption network. To do so, we apply each of the 11 methods to a given year of the network for obtaining possible missing links. Next, we rank the obtained predictions according to the link predictor value and select the top-10 predictions. The accuracy of these predictions is evaluated by verifying the number of missing links that actually appear in future iterations of the network. We have applied this procedure for the 11 methods from 2005 to 2013, and the aggregated fraction of correct predictions is shown in Fig.~\ref{fig:5}A. We have further compared these methods with a random approach where missing links are chosen purely by chance. The accuracy of the random approach was estimated in the same manner and bootstrapped over one hundred realizations for obtaining the confidence interval (results are also shown in Fig.~\ref{fig:5}A). We observe that the methods SimRank, rooted PageRank, and resource allocation have the same accuracy: slightly more than 1/4 of the top-10 links predicted by these methods actually have appeared in future stages of the corruption network. The methods based on the Jaccard similarity, cosine similarity, and association strength have accuracy around 13\%, while the degree product method (also known as preferential attachment method) has an accuracy of 8\%. For the other four methods (Adamic/Adar, common neighbors, HRV, and Katz), none of the top-10 links predicted have appeared in the corruption network. The random approach has an accuracy of 0.2\% with the 95\% confidence interval ranging from 0\% to 5.5\%. Thus, the top seven best-performing methods have a statistically significant predicting power when compared with the random approach. It should also be noted that some of the predicted missing links can eventually appear in future stages of corruption network, that is, by considering scandals occurring after 2014 or even in scandals not yet uncovered. Because of that, the accuracy of methods is likely to be underestimated. In spite of that, all approaches display a high fraction of false positives, that is, nearly 3/4 of top-10 predicted links do not appear on the network.

\section*{Conclusions}
We have studied the dynamical structure of political corruption networks over a time span of 27 years, using as the basis 65 important and well-documented corruption cases in Brazil that together involved 404 individuals.
This unique dataset allowed us to obtain fascinating details about individual involvement in particular scandals, and to analyse key aspects of corruption organization and evolution over the years.

Not surprisingly, we have observed that people engaged in corruption scandals usually act in small groups, probably because large-scale corruption is neither easy to manage nor easy to run clandestine. We have also observed that the time series of the yearly number of people involved in corruption cases has a periodic component with a four-year period that corresponds well with the four-year election cycle in Brazil. This of course leads to suspect that general elections not only reshuffle the political elite, but also introduce new people to power that may soon exploit it in unfair ways.

Moreover, we have shown that the modular structure of the corruption network is indicative of tight links between different scandals, to the point where some could be merged together and considered as one due to their participants having very intricate relationships with one another. By employing the network cartography approach, we have also identified different roles individuals played over the years, and we were able to identify those that are arguably the central nodes of the network. By studying the evolution of the corruption network, we have shown that the characteristic exponential degree distribution exhibits two abrupt variations in years that are associated with key changes in the political power governing Brazil. We further observed that the growth of the corruption network is characterized by abrupt changes in the size of the largest connected component, which is due to the coalescence of different network modules.

Lastly, we have shown that algorithms for predicting missing links can successfully forecast future links between individuals in the corruption network. Our results thus not only fundamentally improve our understanding of the dynamical structure of political corruption networks, but also allow predicting partners in future corruption scandals.

\section*{Author Contributions}
H.V.R., L.G.A.A., and M.P. designed research; H.V.R., L.G.A.A, A.F.M., and E.K.L. performed research; H.V.R., L.G.A.A, and M.P. analyzed data; and H.V.R., L.G.A.A, A.F.M., E.K.L., and M.P. wrote the paper.

\section*{Funding}
This research was supported by CNPq, CAPES (Grants Nos. 440650/2014-3 and 303642/2014-9),  FAPESP (Grant No. 2016/16987-7), and by the Slovenian Research Agency (Grants Nos. J1-7009 and P5-0027).

\end{document}